\documentclass[dvips]{acta}
\usepackage{supertabular,lscape,epsfig}
\usepackage{amssymb}
\usepackage{amsmath}

\usepackage{subfigure}
\usepackage{graphicx}
\usepackage{latexsym}
\usepackage{mathrsfs}
\usepackage{ulem}

\SetPages{0}{0}

\SetVol{99}{9999}

\advance \voffset by 1.00cm\relax
\def\msun{{\,M_{\odot}}}

\def\nuu{\nu_{\rm U}}
\def\nul{\nu_{\rm L}}

\newcommand{\be}{\begin{equation}}
\newcommand{\ee}{\end{equation}}
\begin{document}
\begin{Titlepage}
\Title{Radiative corrections to the neutron star mass inferred
from QPO frequencies}
%
%
%
%
\Author{A~b~r~a~m~o~w~i~c~z$^{1,2,3,4}$, M.A.
\dots
K~l~u~{\'z}~n~i~a~k$^{2,4}$, W.
\dots
Y~u$^4$, W.}
{$^1$Department of Physics, G\"oteborg University, SE-412-96 G\"oteborg,
Sweden \\
e-mail: marek.abramowicz@physics.gu.se\\
$^2$Nicolaus Copernicus Astronomical Center, Polish Academy of
Sciences, ul. Bartycka 18, PL-00-716 Warszawa, Poland\\
e-mail: wlodek@camk.edu.pl\\
$^3$Institute of Physics, Faculty of Philosophy and Science,
Silesian University in Opava, Bezru{\v c}ovo n{\'a}m. 13, CZ-746-01
Opava, Czech Republic\\
$^4$Shanghai Astronomical Observatory, Chinese Academy of
Sciences, 80 Nandan Road, Shanghai 200030, China\\
e-mail: wenfei@shao.ac.cn}
\Received{Month Day, Year}
\end{Titlepage}
\Abstract{The frequencies of kHz QPOs are widely interpreted as
being indicative of the values of characteristic frequencies
related to orbital motion around neutron stars, e.g., the radial
epicyclic frequency. In regions directly exposed to the radiation
from the luminous neutron star these frequencies change with the
luminosity. Including radiative corrections will change the
neutron star mass value inferred from the QPO frequencies.
Radiative forces may also be behind the puzzling phenomenon of
parallel tracks.}{Stars: neutron --- accretion disks --- X-rays:
binaries --- QPOs}
\section{Introduction}
``Twin peak'' QPOs with a pair of physically connected frequencies
$\nuu$, $\nul$ are observed in several X-ray binaries (e.g., van
der Klis 2000). Physical models involve characteristic frequencies
related to orbital motion, either directly (e.g., Morsink \&
Stella 1999), or indirectly through their effect on disk
oscillation properties (e.g., Wagoner 1999, Kato 2001). The values
of these frequencies depend on the mass of the central neutron
star or black hole.

The frequency ratio in all black hole sources in which the twin
peak QPOs have been detected is $\nuu/\nul = 3/2$ exactly. In
several neutron star X-ray binaries the ratio has approximately
the same value, $\nuu/\nul \approx 3/2$. Klu{\'z}niak \&
Abramowicz (2000), Abramowicz \& Klu{\'z}niak (2001), and
Klu{\'z}niak \& Abramowicz (2002), proposed that the twin peak
QPOs may be explained by a non-linear 3:2 resonance in the
epicyclic oscillations of accretion disks around the compact
object in these sources. The
{resonance occurs because of properties of strong
gravity in general relativity and hence}
the model predicts that the QPO
frequencies scale inversely with the {gravitational}
mass of the object,
\begin{equation}
\nu \sim \frac{1}{M}.
\label{inverse-proportion}
\end{equation}
However, Abramowicz, Bulik, Bursa \& Kluzniak (2003) pointed out
that when this model is applied to the atoll and Z sources, where
$\nuu \approx 900\,$Hz, a rather low value for the neutron star
mass is obtained $M \le 1.2\,M_{\odot}$. In the case of Sco X-1
those authors found that the epicyclic resonance model predicts
the mass to be $M = 1\,M_{\odot}$. A similar problem occurs for
most of the neutron star sources, as exhaustively discussed in a
recent paper by Urbanec et al. (2010). Note that the black hole
QPOs have lower frequencies, and accordingly the implied masses
can be much higher, and  for particular values of the Kerr
parameter can be made to correspond to the measured black hole
masses.

Discussing this problem, Urbanec et al. (2010) concluded that the
epicyclic resonance model {\it cannot} explain the QPO phenomenon
in neutron star sources, {\it unless} the oscillating accretion
disks around neutron stars experience some ``fairly non-geodesic''
effects, i.e., when there are strong non-gravitational forces
acting on the disk matter.

In this Note we suggest a natural origin for such ``fairly
non-geodesic'' effects. We expand on an idea already suggested by
Yu (2008, 2011)
{that radiation force affects the orbital frequencies
in QPO sources}, and show {(in Section 3)}
that when the illumination of the
accretion disk by the radiation emitted by the neutron star is
properly included, the 3:2 epicyclic resonance model implies
higher neutron star masses. 
{The main point is that radiation
forces reduce the effective gravity.}
The observed frequencies $\nuu \approx
1\,$kHz may then be consistent with the canonical mass value $M_*
\sim 1.4 M_{\odot}$, and even higher ones.
{We note that Orosz \& Kuulkers (1999) measured a mass of
$1.78\pm 0.23 M_{\odot}$ for the neutron star Cyg X-2, which is known to be
a high luminosity twin kHz QPO source (Wijnands et al. 1998).}
\section{Illuminated epicyclic oscillator}
\begin{figure}[t]
\begin{centering}
\includegraphics[width=0.98\textwidth]{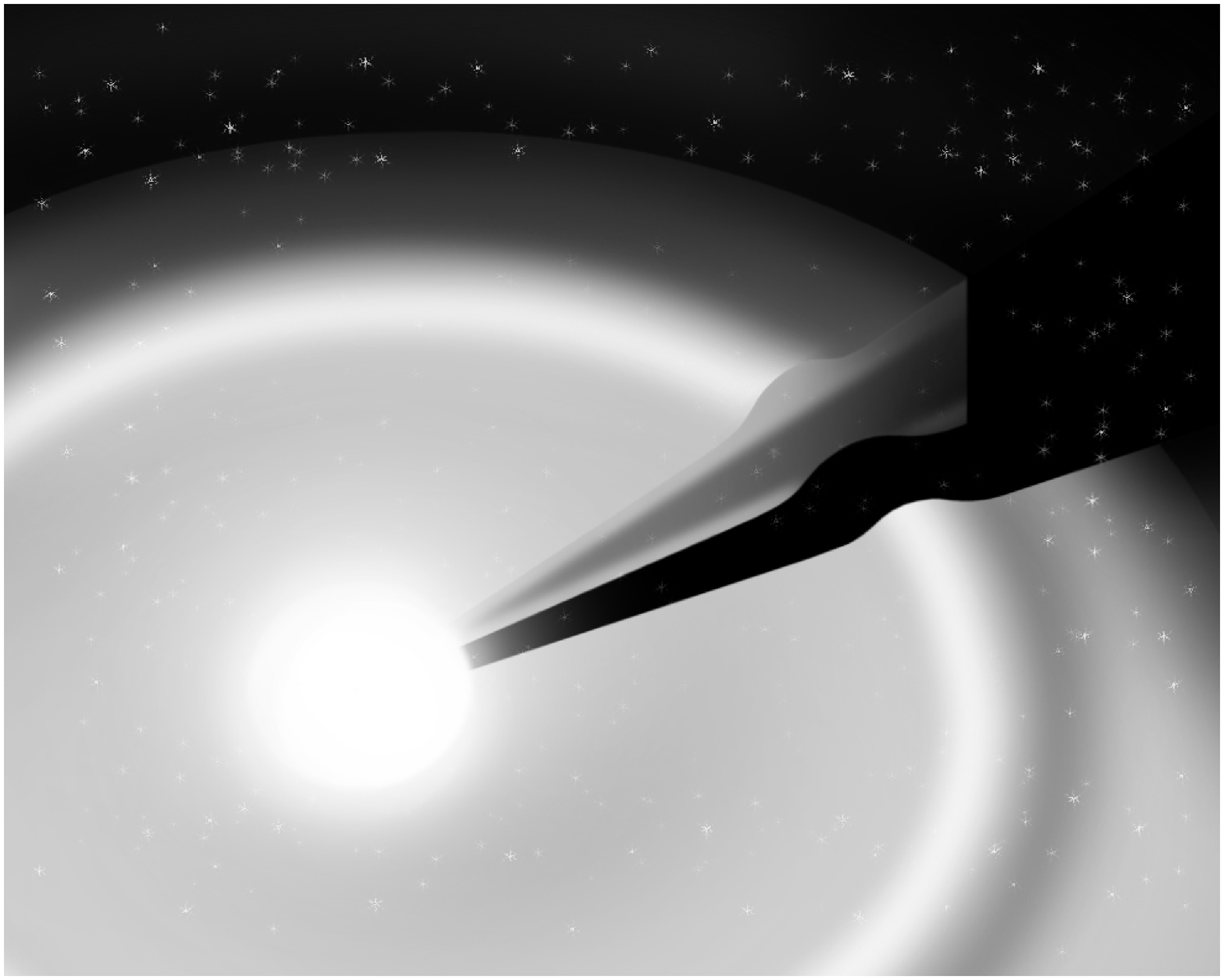}
\end{centering}
\FigCap{Oscillating accretion disk illuminated by the neutron star
boundary layer. The inner disk is optically thin in the vertical direction,
and optically thick in the radial direction (note the shadow in
the cut-away). In the figure, the disk has been truncated for clarity. }
\label{figure-by-malgosia}
\end{figure}
Figure \ref{figure-by-malgosia} illustrates the effect discussed
here, for the epicyclic resonance model.
The interior of a geometrically thin accretion disks around
a neutron star is affected by radiation of the luminous boundary
layer only in a very small region near the inner edge $r_{\rm
in}$. This is because the optical depth in the radial direction is
very large, and the disk is self-shadowing. This has two obvious
consequences. Firstly, the radiative drag (Poynting-Robertson effect)
discussed by Walker \& M\'esz\'aros (1989), and others,
cannot affect motion of matter inside the disk at the location of
the 3:2 resonance radius $r_{3:2} \approx 2r_{\rm in}$. Secondly, the
vertically oscillating part of the disk (shown in Figure
\ref{figure-by-malgosia} as a toroidal ring) periodically emerges
from the disk shadow, and hence, is illuminated by the boundary
layer. If the oscillating torus is optically thin (as will be the case
when the disk is optically thin in the vertical direction) the emergent
ring will briefly be subjected to a radiative force proportional to
the mass of the ring that is momentarily outside the shadow of the disk.
Thus, the effective gravity on the ring is reduced by a factor
proportional to the luminosity of the boundary layer.

In passing, we note that the 
radiation force  caused by this illumination is a function of the
vertical displacement, ${\cal F} = {\cal F}(\delta z)$.
This leads to a coupling between the vertical and radial oscillations
that can be described by a forced oscillator equation,
%
\begin{equation}
\delta {\ddot r} + \omega_r^2 \delta r = {\cal F}_r ( \vert \delta z \vert),
\label{vertical-forced}
\end{equation}
where $\omega_r$ is the radial epicyclic
frequency corresponding to pure gravity (absence of illumination), and
the form of the forcing function is to be calculated
from the exact geometry of illumination.
Such a coupling is a desirable feature in the QPO resonance model.
\section{Corrections to the mass estimated from the oscillatory frequencies}
Observations of the response of the twin kHz QPOs to the flux of
the normal branch oscillations (NBOs) in the manner of an
anti-correlation between the upper kHz QPO frequency and the NBO
flux in Sco X-1 provided  evidence that radiation force affects
the frequency of the orbital signal on sub-second time scales (Yu,
van der Klis \& Jonker 2001). More solid evidence is found for
this anti-correlation in the atoll source 4U 1608-52 (Yu \& van
der Klis 2002). Yu (2008 and 2011)  suggested that the radiation
force can cause the apparent QPO coherence difference between Z
and atoll sources and the apparent coherence drop in several
well-observed atoll sources (Barret et al. 2005, 2008), and that
the radiation force effect can be used to constrain neutron star
masses and radii if both radiation force and orbital frequency can
be measured. { For a clear positive correlation between luminosity
and QPO frequency see e.g., Fig. 1 of van der Klis (2001).}

The radiation force effects that are suggested by
the {correlated behavior} of the kHz QPO frequency  and the
X-ray flux in both atoll sources and Z sources provide a natural
origin of the ``fairly non-geodesic'' effects invoked by Urbanec
et al. (2010). This is because radiation leads to a reduced
``effective gravity'' and consequently the mass $M$ that appears
in (\ref{inverse-proportion}) is not identical with the true
{gravitational} mass
of the star $M_*$, but instead it is the {\it smaller} ``effective
mass''
\begin{equation}
M = M_*\left( 1 - \epsilon \frac{L_*}{L_{\rm Edd}}\right).
\label{effective-mass}
\end{equation}
Here, $L_*$ is that part of the total observed luminosity $L$ that
influences the disk oscillations, and the part that does not will
be $L_i = L - L_*$, while $\,L_{\rm Edd}$ is the Eddington
luminosity, and $\epsilon < 1$ is a factor that depends on details
of the radiation field geometry and of the radiation transfer in
the accretion disk. Accordingly, the mass of the neutron star
predicted by the epicyclic model when radiative corrections are
included is,
\begin{equation}
\frac{M_*}{M_{\odot}} = \left[\frac{f_{3:2}}{\nuu}\right] \left[ 1
- \frac{\epsilon (L - L_i)}{L_{\rm Edd}}\right]^{-1}.
\label{main-formula}
\end{equation}
Here $f_{3:2} \approx 900\,{\rm Hz}$ for the 3:2 resonance in
Schwarzschild metric. For different orbital resonances (discussed
by several authors) the value of this coefficient will be
different.

One may estimate values of $\epsilon$ and $L_i$ from observations.
From (\ref{main-formula}) it follows, in the case of small
$\epsilon L_*$, that
\begin{equation}
{\nuu} = f_{3:2} \left( \frac{\epsilon\msun}{L_{\rm Edd}M_*
}\right)\,L +f_{3:2} \frac{\msun}{M_*}\left( 1 - \frac{\epsilon
L_i}{L_{\rm Edd}}\right).
\label{parallel-tracks}
\end{equation}
If the values of $\epsilon$ and $L_i$  do not change much during
the observation period, and $L_i$ changes from one observation to
another (indicated by the $i$ label), then 
$\nu = a L + b_i$. In this case one may estimate $\epsilon$ and
$L_i$ from the observable parameters $a = a(\epsilon,M_*)$, and
$b_i = b_i(\epsilon,M_*, L_i )$. Thus, Eq.~(\ref{parallel-tracks})
may explain the observed parallel tracks (M\'endez et al. 1999),
{in which the QPO frequency is proportional to the luminosity
on short timescales, but for successive observations
the lines on a luminosity-frequency plot
are displaced in frequency by different offsets.}

%
\section{Conclusions}
In conclusion we want to stress three points:
\begin{itemize}
\item The radiation force from the luminous neutron star will
affect the orbital, and related frequencies,  most strongly at the
inner disk edge.
\item In all previous estimates of the neutron star mass based on
the 3:2 epicyclic resonance QPO model it was assumed that
$\epsilon = 0$, and this implied masses that were too low. The
inclusion of radiation force effects will lead in a natural manner
to an increased mass prediction.  The brighter the neutron star
source is, the more the inferred mass will be raised, since the
mass was underestimated in the original models by an amount
proportional to the strength of the radiation force effect.  For
Sco X-1, if $\epsilon L = 0.5\, L_{\rm Edd}$, the mass estimate would double,
yielding a mass up to 2 solar masses. For atoll sources, the mass
estimate would be raised by 10-20\%, e.g., by $0.1-0.4 \msun$,
assuming $L$ is about 10-20\% $L_{\rm Edd}$.
\item The inclusion of the radiation force may provide an
explanation for the parallel tracks phenomenon.
\end{itemize}
%
%

\Acknow{Research supported in part by the Polish Ministry of Science
(grant NN203 381436). WK acknowledges the hospitality of the Shanghai
Astronomical Observatory. MAA acknowledges the support of the Czech grant MSM
4781305903}


\begin{references}
\refitem{Abramowicz, M.A., \& Klu\'zniak, W.}{2001}
{A\&A}{374}{L19}

\refitem{Abramowicz, M.A., Bulik, T., Bursa, M., Klu\'zniak}{2003}
{A\&A}{404}{L21}

\refitem{Barret, D., Klu\'zniak, W., Olive, J. F., Paltani, S.,
Skinner, G. K.}{2005} {MNRAS}{357}{1288}

\refitem{Barret, D., Boutelier, M.}{2008} {New Astronomy
Reviews}{51}{835}

\refitem{Kato, S.}{2001} {PASJ}{53}{1}

\refitem{Klu\'zniak, W., Abramowicz, M.A.}{2000}{eprint
arXiv:astro-ph/0105057} {}{}

\refitem{Klu\'zniak, W., Abramowicz, M.A.}{2002}{eprint
arXiv:astro-ph/0203314} {}{}

\refitem{M\'endez, M., van der Klis, M., Ford, E. C., Wijnands,
R., van Paradijs, J.}{1999}{ApJ}{511}{L49}

\refitem{Morsink, S.M., Stella, L.}{1999}{ApJ}{513}{827}

\refitem{Orosz, J.E., Kuulkers, E.}{1999}{MNRAS}{305}{132}

\refitem{Urbanec, M., T\"or\"ok, G., {\v S}r\'amkov\'a, E., {\v
C}ech, P., Stuchl\'ik, Z., Bakala, P.}{2010}{A\&A}{522}{72}

\refitem{van der Klis, M.}{2000}{Annual Review of Astronomy and
Astrophysics}{38}{717}

\refitem{van der Klis, M.}{2000}{ApJ}{561}{943}

\refitem{Wagoner, R.V.}{1999}{Phys. Rev.}{311}{259}

\refitem{Wijnands, R., Homan, J., van der Klis, M.
et al.}{1998}{ApJ}{493}{L87}

\refitem{Yu, W., van der Klis, M. \& Jonker,
W.}{2001}{ApJ}{559}{L29}

\refitem{Yu, W. \& van der Klis M.}{2002}{ApJ}{567}{L67}

\refitem{Yu, W.}{2008}{AIP Conference Proceedings}{968}{215}

\refitem{Yu, W.}{2011}{in preparation} {}{}

\refitem{Walker \& M\'esz\'aros}{1989}{ApJ}{346}{844}
\end{references}
\end{document}